\documentclass[conference]{IEEEtran}
\IEEEoverridecommandlockouts
\usepackage{cite}
\usepackage{amsmath,amssymb,amsfonts}
\usepackage{graphicx}
\usepackage{textcomp}
\usepackage{xcolor}

\usepackage{booktabs}
\usepackage{caption}
\usepackage{subcaption}
\usepackage{multirow}
\usepackage{algorithm}

\usepackage[english]{babel}
\usepackage[utf8]{inputenc}
\usepackage[noend]{algpseudocode}

\DeclareMathOperator*{\argmin}{argmin} 

\def\BibTeX{{\rm B\kern-.05em{\sc i\kern-.025em b}\kern-.08em
    T\kern-.1667em\lower.7ex\hbox{E}\kern-.125emX}}
\begin{document}

\title{Representation Learning with Function Call Graph
Transformations for Malware Open Set Recognition
}


\author{\IEEEauthorblockN{Jingyun Jia}
\IEEEauthorblockA{\textit{Department of Computer Engineering and Sciences} \\
\textit{Florida Institute of Technology}\\
Melbourne, FL 32901, US \\
Email: jiaj2018@my.fit.edu}
\and
\IEEEauthorblockN{Philip K. Chan}
\IEEEauthorblockA{\textit{Department of Computer Engineering and Sciences} \\
\textit{Florida Institute of Technology}\\
Melbourne, FL 32901, US \\
Email: pkc@cs.fit.edu}
}

\maketitle

\begin{abstract}
Open set recognition (OSR) problem has been a challenge in many machine learning (ML) applications, such as security. As new/unknown malware families occur regularly, it is difficult to exhaust samples that cover all the classes for the training process in ML systems. An advanced malware classification system should classify the known classes correctly while sensitive to the unknown class. In this paper, we introduce a self-supervised pre-training approach for the OSR problem in malware classification. We propose two transformations for the function call graph (FCG) based malware representations to facilitate the pretext task. Also, we present a statistical thresholding approach to find the optimal threshold for the unknown class. Moreover, the experiment results indicate that our proposed pre-training process can improve different performances of different downstream loss functions for the OSR problem.\smallbreak
\end{abstract}


\section{Introduction}
As machine learning has achieved great success in various domains, there is still a wide range of challenges in the real world. For example, from the security scenario, new malware classes emerge daily. A robust machine learning system for malware detection should be able to classify the known malware classes and recognize the newly unknown malware classes, which is referred as Open Set Recognition (OSR) problem \cite{bendale2016towards}. The OSR problem aims to classify the multiple known classes for a multinomial classification problem while identifying the unknown classes.

In this paper, we follow a two-stage learning approach to address the OSR problem in malware classification. Given the malware assembly files, we first extract the function call graphs (FCGs) as the input representations of the malware samples. Next, to learn better representations for the malware samples, we use a self-supervised pre-training approach for the extracted FCGs. As the self-supervised learning approach needs a pretext task, we propose two transformations for the FCG inputs. Then both original and transformed FCGs are processed by a detransformation autoencoder  (DTAE)  \cite{DBLP:journals/corr/abs-2105-13557}. DTAE involves an encoder and a decoder. The encoder learns the representations for the inputs while the decoder reconstructs the transformed inputs back to their original forms. After pre-training and fine-tuning the representations, we apply a statistical thresholding approach to find the optimal threshold for the OSR tasks. Our contributions include, first, we summarize the characteristics of the malware FCGs. Second, we propose two transformation methods for the malware FCGs to facilitate the self-supervised pre-training process for the OSR tasks. Third, we introduce a statistical thresholding approach for the OSR task, which performs similarly to the manually selected threshold. Finally, our experiments on two different malware datasets indicate that our proposed self-supervised pre-training approach improves the model performance on the OSR tasks.

We organize this paper as follows. In section 2, we review some related research works. In section 3, we first present our proposed approach to the self-supervised pre-training for the malware FCGs, then introduce a statistical thresholding approach for the OSR tasks. Finally, section 4 evaluates the proposed approach through experiment setup and results from the analysis.

\section{Related Work}

\noindent \textbf{Function Call Graphs} The graph features can preserve the structural information between different entities, and have been widely used in many research fields, such as social network recommendation \cite{DBLP:conf/www/Fan0LHZTY19}, molecules structure study \cite{DBLP:conf/icml/GilmerSRVD17} and malware classification \cite{DBLP:conf/codaspy/HassenC17}. Specifically, the researches in \cite{DBLP:journals/tse/Ryder79} and \cite{DBLP:conf/codaspy/HassenC17} extract function call graphs (FCGs) from disassembled binary files. An FCG is a directed graph where the vertices represent the function clusters (procedures), and the edges represent the caller-callee relation between the functions (vertices). As the FCGs have a good performance in saving the interaction information between functions, in this work, we also use malware FCGs as input features for the open set recognition (OSR) problem.
 
 \noindent \textbf{Open Set Recognition} The objective of the OSR problem is to classify the multiple known classes for a multinomial classification problem while identifying the unknown classes. As new and unknown malware class occurs regularly, it is impossible to collect samples that exhaust all the malware classes. An advanced malware classification system should adapt to the open set scenario, classifying the known classes while recognizing the unknown class. Recent work have brought neural network-based approach for the OSR problem such as the works in \cite{bendale2016towards}, \cite{DBLP:conf/sdm/HassenC20} and \cite{DBLP:conf/icann/JiaC21}. OpenMax \cite{bendale2016towards} adapts Extreme Value Theory (EVT) meta-recognition calibration in the penultimate layer of the networks. Further, it redistributes values of the activation vector to estimate the probability of an input being from an unknown class. Hassen and Chan propose ii loss for open set recognition \cite{DBLP:conf/sdm/HassenC20}. It first finds the representations for the known classes during training and then recognizes an instance as unknown if it does not belong to any known classes. MMF \cite{DBLP:conf/icann/JiaC21} is an extension to different types of loss functions (classification loss and representation loss) to facilitate the OSR task. It further separates the known and unknown representations by increasing the signature feature magnitudes of the known classes. Here, we propose a self-supervised learning approach for the malware OSR problem. Adding such a self-supervised pre-training process makes classification loss and representation loss functions more sensitive to the unknown class.
 
 \noindent \textbf{Self-supervised Learning} Self-supervised learning uses a pretext task that is different from the primary task to learn the representations. The pretext task includes autoencoding, classifying transformations such as rotations \cite{DBLP:conf/iclr/GidarisSK18}, intra-sample vs inter-sample transformations in contrastive loss \cite{DBLP:conf/icml/ChenK0H20}, redundancy reduction in learned features from transformations \cite{DBLP:conf/icml/ZbontarJMLD21}. In addition to image recognition applications, more recent research has extended self-supervised learning to graph representation learning. Specifically, Graph contrastive learning (GraphCL) in \cite{DBLP:conf/nips/YouCSCWS20} designs four types of transformations for a graph contrastive learning framework: node dropping, edge perturbation, attribute masking, and subgraph sampling. The experimental results indicate that the beneficial graph transformation technique is dataset-specific. Moreover, Pairwise Half-graph Discrimination (PHD) in \cite{DBLP:conf/ijcai/JinZL00P21} proposes self-supervised multi-scale contrastive learning for graph representation learning. The approach first generates two augmented views based on local and global perspectives from the input graph. Then, the objective function maximizes the agreement between node representations across different views and networks. However, as we will discuss later in Section \ref{sec: fcg-char},  FCGs are sparser than most graph datasets, such as social networks. The existing graph transformation techniques like node dropping and subgraph sampling are less applicable to FCGs. Our work here introduces two different transformations for the malware FCGs inputs. And then, we adopt the same learning strategy as in DTAE \cite{DBLP:journals/corr/abs-2105-13557}, i.e., reconstructing the transformed inputs back to original forms to improve the quality of learned representations.


\section{Approach}
\label{sec: approach}
The objective of open set recognition (OSR) is to classify the known classes and the unknown classes even when the collected training samples cannot exhaust all the classes. An advanced malware classification system that utilizes OSR techniques can classify the known malware families while identifying the unknown malware family. Hassen and Chan \cite{DBLP:conf/codaspy/HassenC17} convert malware assembly files to FCGs as OSR input. Here, we also use the FCGs as input samples. To learn better representations for the OSR problem in malware classification, we introduce a self-supervised pre-training process to learn low-level features of the malware samples. Based on the FCGs characteristics, we propose two transformation methods for malware FCGs to facilitate the pretext task. Moreover, we introduce a statistical method to identify unknown instances.

\subsection{Malware Function Call Graphs (FCGs)}
\label{sec:fcg}
Previous research works have proposed various ways to extract features for malware classifications: Schultz et al. \cite{DBLP:conf/sp/SchultzEZS01} extract features from printable strings in malware binaries. Hu et al. \cite{DBLP:conf/usenix/HuSBG13} extract features from instruction n-grams. Hassen and Chan \cite{DBLP:conf/codaspy/HassenC17} convert malware assembly files to FCGs as input features. The FCGs can better preserve structural information between functions. Thus, in this paper, we adopt the same FCGs as in \cite{DBLP:conf/codaspy/HassenC17}. The system first extracts FCG representations from dissembled binaries. In the FCGs, the vertices are functions, and edges are the interactions (calls) between functions. Then based on the instruction opcode sequence, it clusters the functions using Locality Sensitive Hashing (LSH), and the vertices (functions) are then arbitrarily labeled with cluster-ids. 

The extracted FCGs are directed graph representations of the dissembled malware binaries, with function clusters as the graph vertices and the caller-callee relations between functions as graph edges. As the cluster ids are arbitrarily assigned, we will get different isomorphic graphs for the same malware binaries when we change the order of the cluster ids.

\begin{table}
\caption{Graph statistics for datasets in function call graphs (FCGs), biochemical molecules (BMs) and social networks (SN). The statistics includes: average number of vertices, average number of degrees and \% of vertices that are neighbors (Degree/Vertex), average number of connected components (C.C.), average size of each connected components and relative connected components size (C.C. Size/Vertex). }
\scalebox{.85}{
\begin{tabular}{| l | c | cccc |}
\hline
Dataset & Category  & Vertex & Degree (/Vertex) & C.C. & C.C. Size (/Vertex)\\ \hline\hline
MS & FCGs & 27.55 & 1.66(6\%) & 14.99 & 3.74(16\%)\\
AG & FCGs & 31.73 & 3.31(10\%) & 16.97 & 2.28(7\%)\\ \hline
MUTAG & BMs & 17.93 & 1.10 (6\%) & 3.49 & 5.86(33\%)\\
PROTEINS & BMs & 39.06 & 1.86(5\%) & 4.75 & 9.78(25\%) \\ 
\hline
COLLAB & SNs & 74.49 & 32.99(44\%) & 4.65 & 30.36(41\%) \\ 
DBLP\_v1 & SNs & 10.48 & 1.87(18\%) & 1.93 & 6.12(58\%) \\
\hline
\end{tabular}}
    \label{tab: stats}
\end{table}

\subsection{FCG characteristics}
\label{sec: fcg-char}
In this subsection, we compare the characteristics of the FCGs of malware datasets with two other categories of graphs: biomedical molecules (BMs) and social networks (SNs) in Table \ref{tab: stats}. Specifically, we compare the FCGs extracted from two malware datasets: Microsoft Challenge (MC) and Android Genome (AG) (see section \ref{sec: exp} for more details) with MUTAG \cite{DBLP:conf/icml/KriegeM12}, PROTEINS \cite{DBLP:conf/ismb/BorgwardtOSVSK05}, COLLAB \cite{DBLP:conf/kdd/YanardagV15} and DBLP\_v1 \cite{grand-lab}. In the table, ``Vertex'' and ``Degree'' are the average numbers of vertices and degrees in each dataset. We also measure the average percentage of vertices that are neighbors by dividing the number of degrees by the number of vertices. Moreover, we calculate the average number of connected components (C.C.) and the average size of connected components (C.C. Size) for each dataset. Also, we divide the size of the C.C. by the number of vertices to measure the relative C.C. Size. Comparing the graph statistics of the FCGs with the other categories, we conclude two characteristics of the FCGs.

First, FCGs are sparser (i.e., have fewer direct neighbors) than the graphs from the other two categories, especially social networks. In the COLLAB dataset, the average degree of a graph is 32.99, which means 44\% of the vertices are direct neighbors. Meanwhile, 6\% of vertices are direct neighbors in the MS dataset and 10\% for the AG dataset. 

Second, FCGs have more and relatively small connected components than the other two categories of graphs. From Table \ref{tab: stats}, both malware FCGs contain around 15 connective components, while the datasets from the other two categories contain less than five connected components. Furthermore, the average sizes of each connected component in the two malware FCGs are less than 4, which means less than four vertices are connected while isolated from the rest of the vertices. Especially for the AG dataset, the connected components are of size 2.28 on average, which is only 7\% of the total vertices. The relative connected components size is above 25\% of the total vertices for the four datasets from the other two categories. Notably, the relative size of connected components in DBLP\_v1 dataset reaches 58\%. 



\begin{figure}[t]
\centering
\begin{subfigure}[t]{0.16\textwidth}
    \begin{subfigure}{\textwidth}
      \renewcommand\thesubfigure{\alph{subfigure}1}
      \centering
      \includegraphics[width=.95\textwidth]{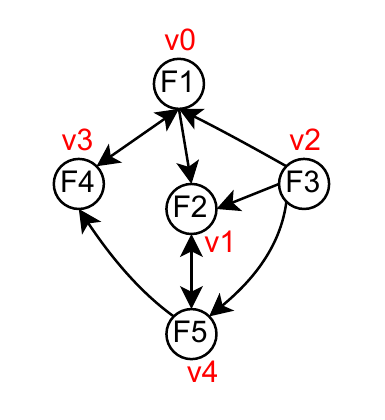}
      \caption{}
      \label{fig: fcg-org-fig}
    \end{subfigure}
    \begin{subfigure}{\textwidth}
      \addtocounter{subfigure}{-1}
      \renewcommand\thesubfigure{\alph{subfigure}2}
      \centering
      \includegraphics[width=.95\textwidth]{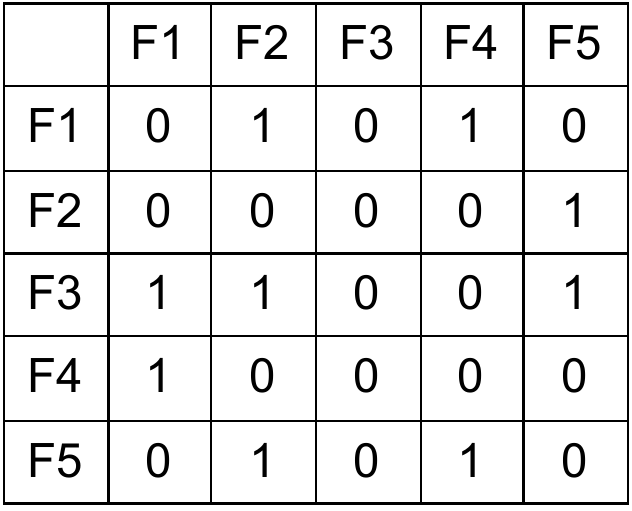}
      \caption{}
      \label{fig:fcg-org-adj}
    \end{subfigure}
    \addtocounter{subfigure}{-1}
    \caption{Original}
    \label{fig: fcg-org}
    \end{subfigure}%
        \begin{subfigure}[t]{0.16\textwidth}
    \begin{subfigure}{\textwidth}
      \renewcommand\thesubfigure{\alph{subfigure}1}
      \centering
      \includegraphics[width=.95\textwidth]{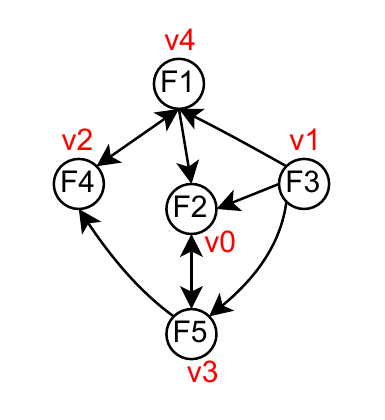}
      \caption{}
      \label{fig: fcg-shift-fig}
    \end{subfigure}
    \begin{subfigure}{\textwidth}
      \addtocounter{subfigure}{-1}
      \renewcommand\thesubfigure{\alph{subfigure}2}
      \centering
      \includegraphics[width=.95\textwidth]{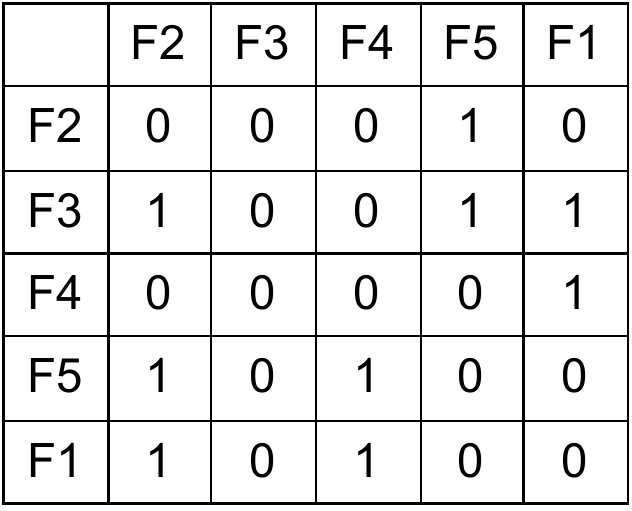}
      \caption{}
      \label{fig: fcg-shift-adj}
    \end{subfigure}
    \addtocounter{subfigure}{-1}
    \caption{FCG-shift}
    \label{fig: fcg-shift}
        \end{subfigure}%
          \begin{subfigure}[t]{0.16\textwidth}
    \begin{subfigure}{\textwidth}
      \renewcommand\thesubfigure{\alph{subfigure}1}
      \centering
      \includegraphics[width=.95\textwidth]{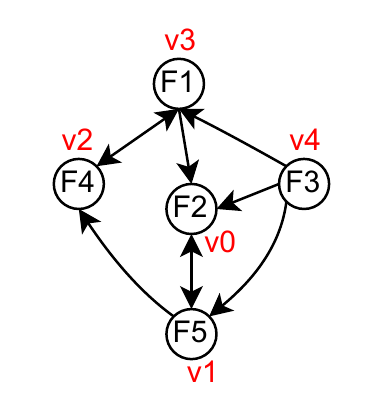}
      \caption{}
      \label{fig: fcg-random-fig}
    \end{subfigure}
    \begin{subfigure}{\textwidth}
      \addtocounter{subfigure}{-1}
      \renewcommand\thesubfigure{\alph{subfigure}2}
      \centering
      \includegraphics[width=.95\textwidth]{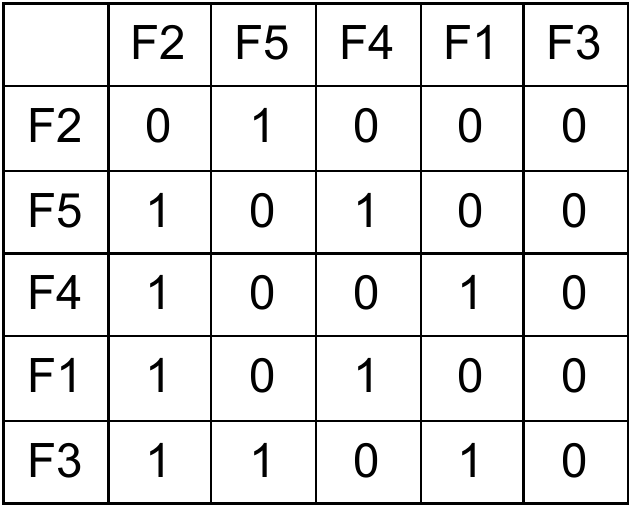}
      \caption{}
      \label{fig: fcg-random-adj}
    \end{subfigure}
    \addtocounter{subfigure}{-1}
    \caption{FCG-random}
        \label{fig: fcg-random}
        \end{subfigure}

\caption{Transformations of FCG adjacency matrix}
\label{fig: fcg-transform}
\end{figure}

\subsection{FCG transformations}
\label{sec: transformations}
Self-supervised learning generally involves input transformations to achieve pretext tasks to learn better representations of input samples. The research in \cite{DBLP:conf/nips/YouCSCWS20} finds that the optimal input transformation method is task-relevant, and it concludes that node dropping and subgraph sampling are generally beneficial across biochemical molecules and social networks datasets. The node dropping transformation creates a new graph view by discarding a specific set of vertices and edges from the original input graph. As the FCGs have fewer direct neighbors and are sparser than other graph datasets, discarding vertices and their edges will remove more neighborhood information. Thus the node dropping transformation is less applicable to the malware FCGs. The subgraph sampling transformation creates a new graph view by sampling a
subgraph from the original input graph via a random walk. From the second characteristic of the FCGs, the FCGs contain more connected components (around 15 for the FCGs dataset from Table \ref{tab: stats}). Since a random walk subgraph sampling will keep one connected component and discard the rest (14 out of 15), the subgraph sampling will discard more than 90\% information. Thus subgraph sampling is not desirable in learning the representations of the FCGs. 

As FCGs can be represented by adjacency matrices, and the ordering of vertices in the matrices is arbitrary. Here, we propose two types of transformations: FCG-shift and FCG-random for the malware FCGs. The two transformations generate a new isomorphic view by altering the ordering of vertices. Given the original order of clusters-ids assignment as Figure \ref{fig: fcg-org}, the FCG-shift transformation randomly select a pivots $n$, and then shift the cluster-ids assignments $n$ positions to the left. For example, in Figure \ref{fig: fcg-shift}, the order of vertices (cluster-ids) is shifted one position to the left. The original vertex order ``F1'', ``F2'', ``F3'', ``F4'', ``F5'' becomes ``F2'', ``F3'', ``F4'', ``F5'', ``F1''. The FCG-random transformation randomly permute the order of vertices and generated new adjacency matrices based on the permuted vertex order. In Figure \ref{fig: fcg-random}, after the random permutation, the original vertex order ``F1'', ``F2'', ``F3'', ``F4'', ``F5'' becomes ``F2'', ``F5'', ``F4'', ``F1'', ``F3''. Both FCG-shift and FCG-random maintain the orignal FCGs' properties without information loss by generating isomorphic graphs to the original graphs.

\subsection{Representation Learning}
In this work, we follow the two-stage learning strategy to learn the representations of input malware FCGs. We adopt the self-supervised learning strategy to initial the network with low-level representations in the first stage. In the second stage, we fine-tune the network with different loss functions to extract the discriminative representations.

\begin{figure*}[t]
 \begin{subfigure}[b]{\textwidth}
 \centering
               \includegraphics[width=0.7\linewidth]{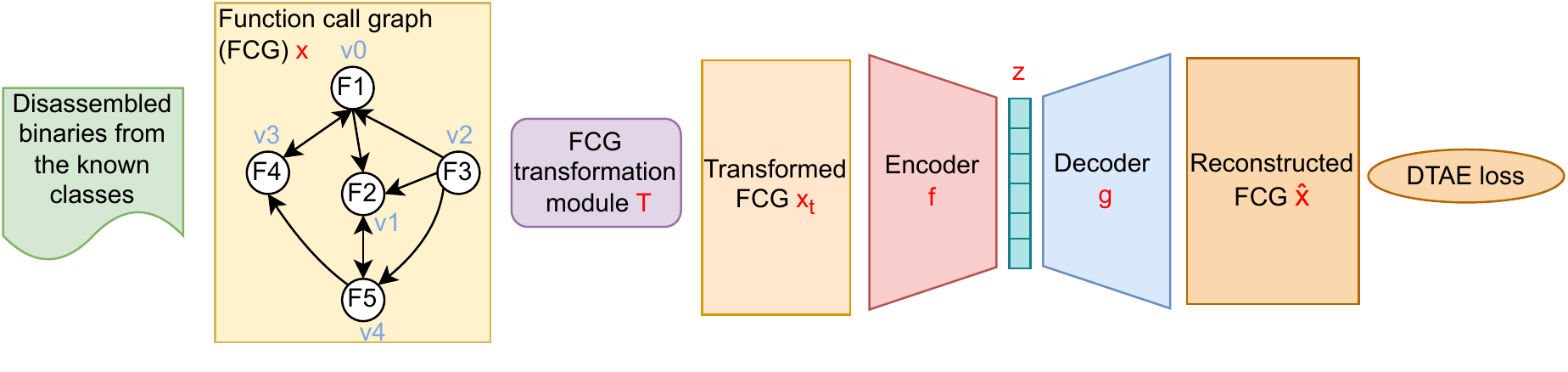}
                \caption{Pre-training step}
                \label{fig: pre-train}
        \end{subfigure} \\
  \begin{subfigure}[b]{0.49\textwidth} 
  \centering
  \includegraphics[width=0.95\linewidth]{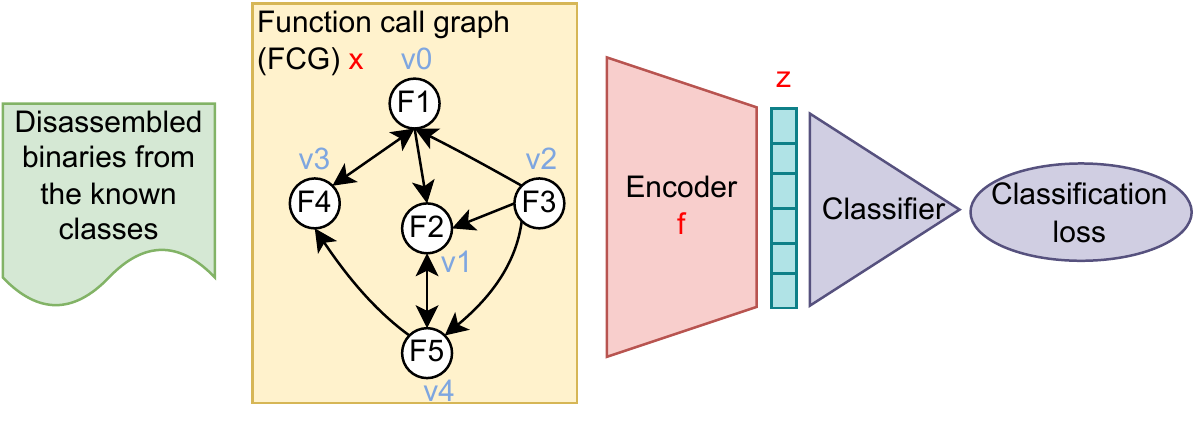}
                \caption{Fine-tuning step with classification loss}
                \label{fig: fine-tune-cls}
        \end{subfigure}%
  \begin{subfigure}[b]{0.49\textwidth}                
  \centering
  \includegraphics[width=0.8\linewidth]{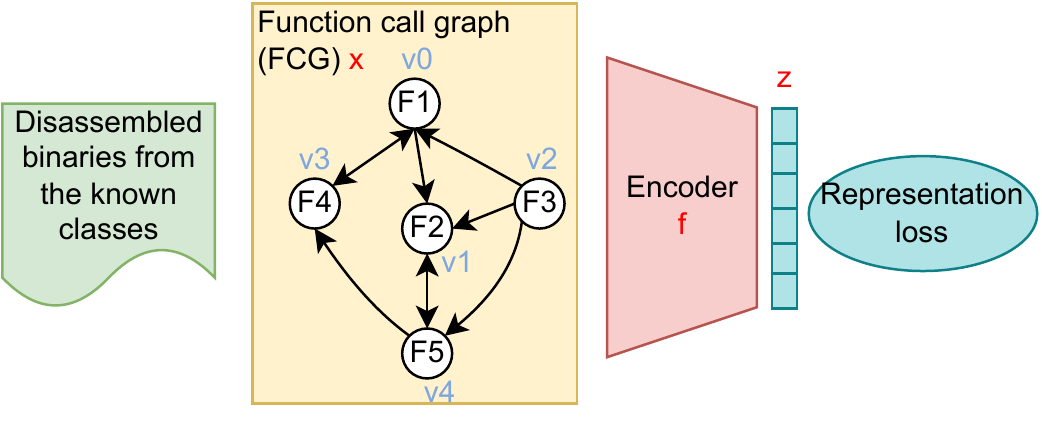}
                \caption{Fine-tuning step with representation loss}
                \label{fig: fine-tune-rep}
        \end{subfigure}%
\caption{The training process of using detransformation autoencoder.}
\label{fig: de-transformation-autoencoder-training}
\end{figure*}


\subsubsection{Pre-training stage}
With the proposed FCG transformations, we adopt detransformation autoencoder (DTAE) proposed in \cite{DBLP:journals/corr/abs-2105-13557} as our pretext task here to pre-train the the network. As depict in Figure \ref{fig: pre-train}, given an input disassembled binaries of the malware samples from the known classes, we first extract its FCG $x_i$. Then the FCG transformation module $T(.)$ transforms the original FCG to its correlated views $x_{it}$. Next, the encoder $f(.)$ learns the representations $z$ of the transformed FCG $x_{it}$, and the decoder reconstruct the representation $z$ back to its original FCG format $\hat{x}_{it}$. Assuming we have $M$ transformations for $N$ FCG inputs. The learning process of neural network-based encoder-decoder structure is guided by DTAE loss:

\begin{equation}
    \mathcal{L}_\text{DTAE} = \frac{1}{2} \sum^{M-1}_{t=0}\sum^N_{i=1}(x_i - \hat{x}_{it})^2 
\label{eq: detrans}
\end{equation}

In this paper, we transform FCGs four times for each experiment, i.e., $M=4, t \in \{0, 1, 2, 3\}$. 


\subsubsection{Fine-tuning stage} After pre-training the neural network with transformed inputs, we fine-tune the encoder and presentation layer (z) with the original inputs for the downstream tasks. Here, we consider two types of loss functions for fine-tuning: classification loss and representation loss. The objective of classification loss is to explicitly lower the training data's classification error in the decision layers, such as cross-entropy loss. When using classification loss as the fine-tuning loss function, we connect the presentation layer with a classifier, which associates with a classification loss function as shown in Figure \ref{fig: fine-tune-cls}. The objective of representation loss functions is to learn better representations of training data. The representation loss functions are normally applied to the representation layers, such as triplet loss. When using representation loss as the fine-tuning loss function, we directly constrain the representation layer with a representation loss function, as shown in Figure \ref{fig: fine-tune-rep}.

\subsection{Open Set Recognition}
\label{sec: osr}
After fine-tuning the encoder with the original FCG inputs, we extract the learned representations $z$ for the malware input. We utilize the distances between the representations for the open set recognition (OSR) task: classifying the known classes and identifying the unknown class. 

For a known class $k$ that participant in the training process, we first find its representation centroid as prototype $\mu_k$. Given the representation $z_i$ for sample $i$ from class $k$ (i.e. $y_i = k$), we can calculate the prototype as:
\begin{equation}
\begin{aligned}
\mu_k = \frac{1}{N_k}\sum_{i=1}^{N_k} z_i,
\end{aligned}
\end{equation}

where $N_k$ is the number of samples in class $k$. After obtaining the prototypes, we introduce a statistical method to perform the OSR task. Specifically, we calculate the mean $m_k$ and standard deviation $s_k$ of the distances $d_i$ from the training samples to the prototype $k$.

\begin{equation}
m_k = \frac{1}{N_k}\sum_{i=1}^{N_k} d_i 
\end{equation}

\begin{equation}
s_k = \sqrt{\frac{\sum_{i=1}^{N_k}(z_i-m_k)^2}{N_k}}
\end{equation}

Then we normalize the distances between representations and prototypes based on the prototypes' means and standard deviations, and calculate the outlier score based on the least of standard deviations to the prototype :

\begin{equation}\label{eq:outlier}
outlier\_score(x) = \min_{1\leq{k}\leq{C}} \frac{ \|D(\mu_k, z) - m_k \|}{s_k},
\end{equation}

where $C$ is the number of known classes, and $z$ is the learned representation of input $x$. $D(.,.)$ is a distance function, and we use euclidean distances in this paper. Based on the Empirical Rule, a test instance can be recognized as ``unknown" if its outlier score is more significant than three standard deviations.

\begin{equation}\label{eq:intra}
    y=
    \begin{cases}
      unknown,& \text{if$\ outlier\_score(x) > 3$} \\
      \argmin\limits_{1\leq{k}\leq{C}} \frac{ \|D(\mu_k, z) - m_k \|}{s_k},& \text{otherwise}
    \end{cases}
\end{equation}

\section{Experiments}
\label{sec: exp}
We evaluate the proposed self-supervised pre-training method with two types of downstream loss functions: triplet loss \cite{DBLP:conf/cvpr/SchroffKP15} (representation loss) and cross-entropy loss (classification loss). Moreover, we test the proposed approach on two malware datasets:

\noindent \textbf{Microsoft Challenge (MC)} \cite{https://doi.org/10.48550/arxiv.1802.10135} contains disassembled malware samples from 9 families:``Ramnit'', ``Lollipop'', ``Kelihos ver3'', ``Vundo'', ``Simda'', ``Tracur'', ``Kelihos ver1'', ``Obfuscator.ACY '' and ``Gatak''. We use 10260 samples that can be correctly parsed then extracted their FCGs as in \cite{DBLP:conf/codaspy/HassenC17} for the experiment. To simulate an open-world dataset, we randomly pick six classes of digits as the known classes participant in the training, while the rest are considered as unknowns that only exist in the test set.

\noindent \textbf{Android Genome (AG)} consists of 1,113 benign android apps and 1,200 malicious android apps. The benign samples are provided by our colleague, and the malicious samples are from \cite{zhou_jiang}. We select nine families with a relatively larger size for the experiment to be fairly split into the training and test sets. The nine families contain 986 samples in total. We first use \cite{DBLP:conf/ccs/GasconYAR13} to extract the function instructions and then generated the FCGs as in \cite{DBLP:conf/codaspy/HassenC17}. Also, to simulate an open-world scenario as the MC dataset, we randomly pick six digits as the known classes in the training set while considering the rest as the unknown class, which are only used in the test phase.

\begin{table*}[t]
\caption{The average AUC scores of 30 runs at 100\% and 10\% FPR of OpenMax and a group of 5 methods for each of the two types of loss functions (ce and triplet): without pre-training, pre-training via DTAE with transformations node dropping (ND), Subgraph sampling (SS), FCG-shift and FCG-random. The values in bold are the highest values in each group. The underlined values show statistically significant improvements (t-test with 95\% confidence) comparing with OpenMax.}
\resizebox{0.95\textwidth}{!}{%
\begin{tabular}{l l  c  c c}
\toprule
 & \multicolumn{1}{c}{}  & \multicolumn{1}{c}{OpenMax}&
 \multicolumn{1}{c}{ce}  & \multicolumn{1}{c}{triplet} \\ \midrule 
  & FPR &   & No pre-training / ND / SS / FCG-shift (ours) / FCG-random (ours) & No pre-training / ND / SS / FCG-shift (ours) / FCG-random (ours) \\ 
\multirow{2}{*}{MC} & 100\% & 0.880{\tiny$\pm$0.037} & 0.918{\tiny$\pm$0.036} / 0.914{\tiny$\pm$0.063} / 0.626{\tiny$\pm$0.054} / \underline{0.938{\tiny$\pm$0.015}} / \underline{\textbf{0.947{\tiny$\pm$0.011}}}  & 0.929{\tiny$\pm$0.020} / 0.919{\tiny$\pm$0.032} / 0.723{\tiny$\pm$0.071} / \underline{0.932{\tiny$\pm$0.017}} / \underline{\textbf{0.933{\tiny$\pm$0.015}}} \\ 
 & 10\% & 0.040{\tiny$\pm$0.003} & 0.053{\tiny$\pm$0.008} / 0.053{\tiny$\pm$0.014} / 0.018{\tiny$\pm$0.005} / \underline{0.061{\tiny$\pm$0.003}} / \underline{\textbf{0.063{\tiny$\pm$0.003}}} & 0.058{\tiny$\pm$0.004} / 0.056{\tiny$\pm$0.006} / 0.036{\tiny$\pm$0.008} / \underline{\textbf{0.061{\tiny$\pm$0.003}}} / 
 \underline{\textbf{0.061{\tiny$\pm$0.003}}}\\  
\multirow{2}{*}{AG} & 100\% & 0.457{\tiny$\pm$0.200} & 0.852{\tiny$\pm$0.056} / 0.820{\tiny$\pm$0.128} / 0.418{\tiny$\pm$0.080} / \underline{\textbf{0.865{\tiny$\pm$0.060}}} / \underline{0.854{\tiny$\pm$0.062}} & 0.868{\tiny$\pm$0.046} / 0.818{\tiny$\pm$0.124} / 0.427{\tiny$\pm$0.094} / \underline{0.873{\tiny$\pm$0.036}} / \underline{\textbf{0.883{\tiny$\pm$0.035}}}\\
 & 10\% & 0.001{\tiny$\pm$0.001} & 0.021{\tiny$\pm$0.012} / 0.019{\tiny$\pm$0.016} / 0.002{\tiny$\pm$0.002} / \underline{\textbf{0.022{\tiny$\pm$0.013}}} / \underline{0.019{\tiny$\pm$0.009}} & 0.024{\tiny$\pm$0.010} / 0.018{\tiny$\pm$0.011} / 0.002{\tiny$\pm$0.002} / \underline{0.025{\tiny$\pm$0.011}} / \underline{\textbf{0.027{\tiny$\pm$0.011}}}\\ \bottomrule

\end{tabular}}
\label{tab:auc}
\end{table*}

\subsection{Experimental Setup}
As described in Section \ref{sec: approach}, our proposed approach first extracts the FCGs from the malware samples, then uses self-supervised DTAE \cite{DBLP:journals/corr/abs-2105-13557} for pre-training before applying downstream fine-tuning tasks. We experiment with classification loss (cross-entropy loss: ce) and representation loss (triplet loss: triplet) as loss functions in the fine-tuning network for the OSR tasks. To demonstrate that our proposed approach is effective for OSR problems, we compare our approach with OpenMax\cite{bendale2016towards}. Moreover, to prove that the self-supervised pre-training step benefits the OSR tasks, we compare the results of using and not using self-supervised pre-training for the two types of loss functions mentioned above. 

As illustrated in Figure \ref{fig: pre-train}, the pre-trained network contains an encoder and a decoder. Furthermore, the learned encoder is fine-tuned with downstream OSR tasks. For the encoder, the padded input layer is of size (67,67) for both MC and AG datasets. Two non-linear convolutional layers follow the input layer with 32 and 64 nodes. We apply the max-pooling layers with kernel size (3, 3) and strides (2, 2) as well as batch normalization after each convolutional layer. After a convolutional block, we add one fully connected non-linear layer with 256 hidden units before the representation layer, containing six dimensions.
Moreover, We use the Relu activation function and set the Dropout's keep probability as 0.2. We use Adam optimizer with a 0.001 learning rate. The decoder in the pre-trained network is simply the reverse of the encoder in our experiments. The encoder and representation layer maintain the same architecture and hyperparameters in the fine-tuning network. Meanwhile, the decoder is replaced with different fully connected layers associated with different loss functions.

\begin{table*}

\caption{The average F1 scores of 30 runs of OpenMax and a group of 6 methods (without pre-training using manually selected threshold as baseline, without pre-training using statistical threshold, pre-training via DTAE with transformations node dropping, subgraph sampling, FCG-shift and FCG-random) for each of the two types of loss functions (ce and triplet). The values in bold are the highest values in each group. The underlined values are statistical significant better than OpenMax.}
\centering
\resizebox{0.76\textwidth}{!}{%
\begin{tabular}{l l ccc ccc}
\toprule
 &  & \multicolumn{3}{c}{MC} & \multicolumn{3}{c}{AG} \\ \midrule
 & & Known & Unknown & Overall & Known & Unknown & Overall \\ \cmidrule(l){3-5} \cmidrule(l){6-8} 
 \multirow{1}{*}{OpenMax} &  & 0.891{\tiny$\pm$0.006} & 0.737{\tiny$\pm$0.010} & 0.869{\tiny$\pm$0.006} & 0.408{\tiny$\pm$0.190} & 0.640{\tiny$\pm$0.163} & 0.441{\tiny$\pm$0.184}  \\ \midrule
\multirow{6}{*}{ce} & No pre-training (manual threshold) & \textbf{0.899{\tiny$\pm$0.010}}  & 0.703{\tiny$\pm$0.061} & 0.871{\tiny$\pm$0.017} & 0.683{\tiny$\pm$0.117} & 0.540{\tiny$\pm$0.329} & 0.663{\tiny$\pm$0.120} \\
& No pre-training (statistical threshold) & 0.890{\tiny$\pm$0.021} & 0.663{\tiny$\pm$0.176} & 0.858{\tiny$\pm$0.042} & 0.705{\tiny$\pm$0.088} & 0.512{\tiny$\pm$0.363} & 0.678{\tiny$\pm$0.120} \\
& Node dropping & 0.852{\tiny$\pm$0.077} & 0.715{\tiny$\pm$0.097} & 0.833{\tiny$\pm$0.078} & 0.684{\tiny$\pm$0.176} & 0.636{\tiny$\pm$0.339} & 0.677{\tiny$\pm$0.181} \\
& Subgraph sampling & 0.000{\tiny$\pm$0.000} & 0.384{\tiny$\pm$0.000} & 0.055{\tiny$\pm$0.000} & 0.006{\tiny$\pm$0.018} & 0.616{\tiny$\pm$0.210} & 0.093{\tiny$\pm$0.016} \\
& FCG-shift (ours) & \underline{0.896{\tiny$\pm$0.010}} & \underline{0.765{\tiny$\pm$0.024}} & \underline{0.878{\tiny$\pm$0.011}} & \underline{\textbf{0.743{\tiny$\pm$0.088}}} & \textbf{0.612{\tiny$\pm$0.327}} & \underline{\textbf{0.724{\tiny$\pm$0.113}}} \\
& FCG-random (ours) & \underline{0.898{\tiny$\pm$0.012}} & \underline{\textbf{0.774{\tiny$\pm$0.025}}} & \underline{\textbf{0.880{\tiny$\pm$0.013}}} & \underline{0.647{\tiny$\pm$0.129}} & 0.608{\tiny$\pm$0.318} & \underline{0.641{\tiny$\pm$0.127}} \\ \midrule
\multirow{6}{*}{triplet} & No pre-training (manual threshold) & 0.905{\tiny$\pm$0.007} & 0.728{\tiny$\pm$0.035} & 0.879{\tiny$\pm$0.011} & 0.753{\tiny$\pm$0.074} & 0.789{\tiny$\pm$0.133} & 0.758{\tiny$\pm$0.068} \\
& No pre-training (statistical threshold) & 0.903{\tiny$\pm$0.010} & 0.749{\tiny$\pm$0.036} & 0.881{\tiny$\pm$0.013} & 0.771{\tiny$\pm$0.059} & \textbf{0.827{\tiny$\pm$0.093}} & 0.779{\tiny$\pm$0.054} \\
& Node dropping & 0.884{\tiny$\pm$0.036} & 0.736{\tiny$\pm$0.046} & 0.862{\tiny$\pm$0.037} & 0.679{\tiny$\pm$0.184} & 0.768{\tiny$\pm$0.170} & 0.692{\tiny$\pm$0.171} \\
& Subgraph sampling & 0.014{\tiny$\pm$0.075} & 0.372{\tiny$\pm$0.069} & 0.065{\tiny$\pm$0.054} & 0.011{\tiny$\pm$0.061} & 0.657{\tiny$\pm$0.135} & 0.104{\tiny$\pm$0.036} \\
& FCG-shift (ours) & \underline{\textbf{0.906{\tiny$\pm$0.007}}} & \underline{0.758{\tiny$\pm$0.021}} & \underline{\textbf{0.885{\tiny$\pm$0.008}}} & \underline{0.745{\tiny$\pm$0.074}} & 0.744{\tiny$\pm$0.250} & \underline{0.745{\tiny$\pm$0.092}} \\
& FCG-random (ours) & \underline{\textbf{0.906{\tiny$\pm$0.007}}} & \underline{\textbf{0.763{\tiny$\pm$0.020}}} & \underline{\textbf{0.885{\tiny$\pm$0.008}}} & \underline{\textbf{0.776{\tiny$\pm$0.061}}} & \underline{0.819{\tiny$\pm$0.166}} & \underline{\textbf{0.782{\tiny$\pm$0.067}}} \\
\bottomrule
\end{tabular}}
\label{tab:f1}
\end{table*}

\begin{figure*}
\centering
\begin{subfigure}[t]{.315\textwidth}
    \includegraphics[trim={0 0 2cm 0},clip, width=\linewidth, height=0.2\textheight]{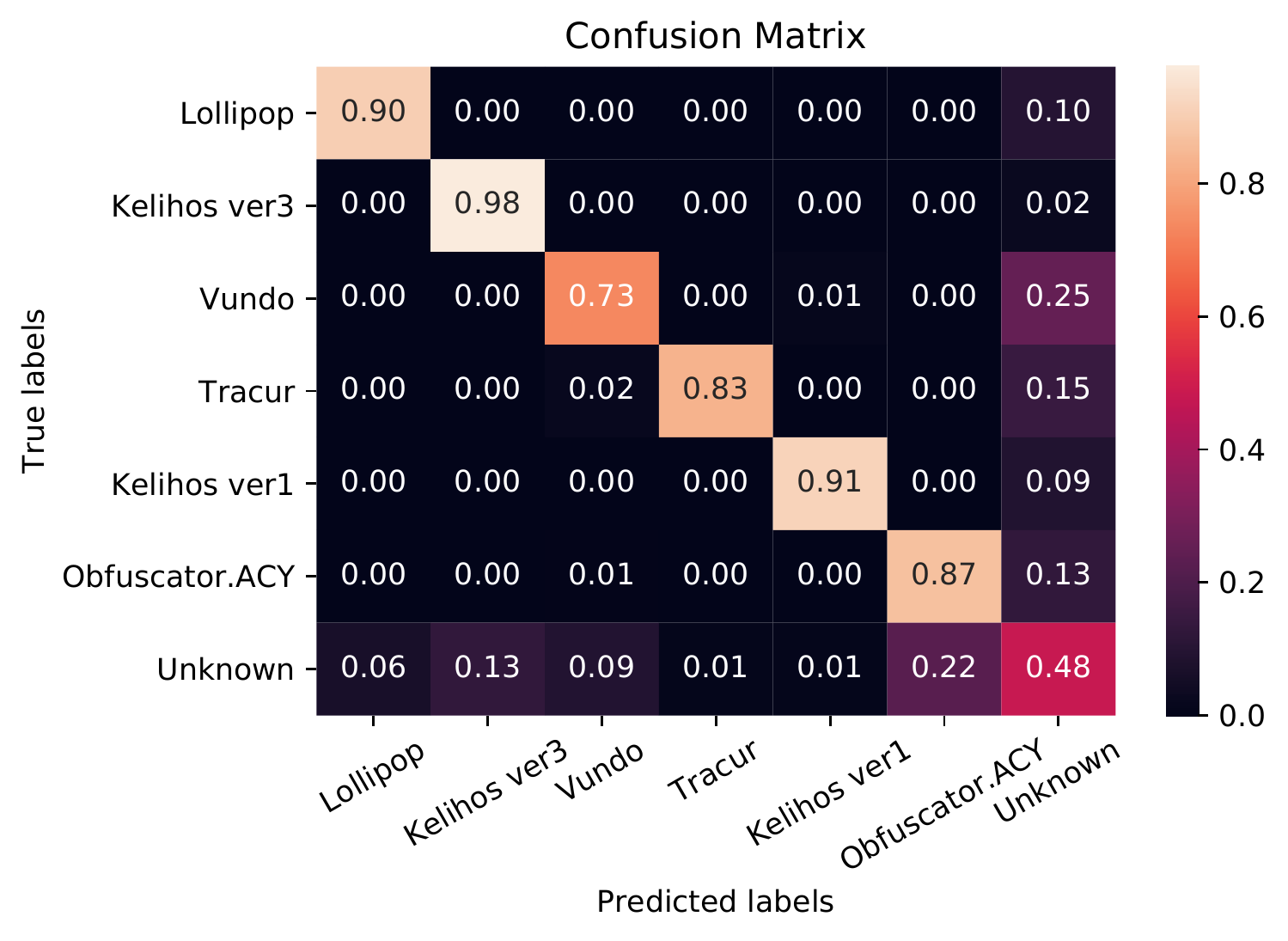}
    \caption{Without pre-training (ce)}
\label{fig: hm1}
\end{subfigure}\hfill
\begin{subfigure}[t]{.315\textwidth}
    \includegraphics[trim={0 0 2cm 0},clip,width=\linewidth, height=0.2\textheight]{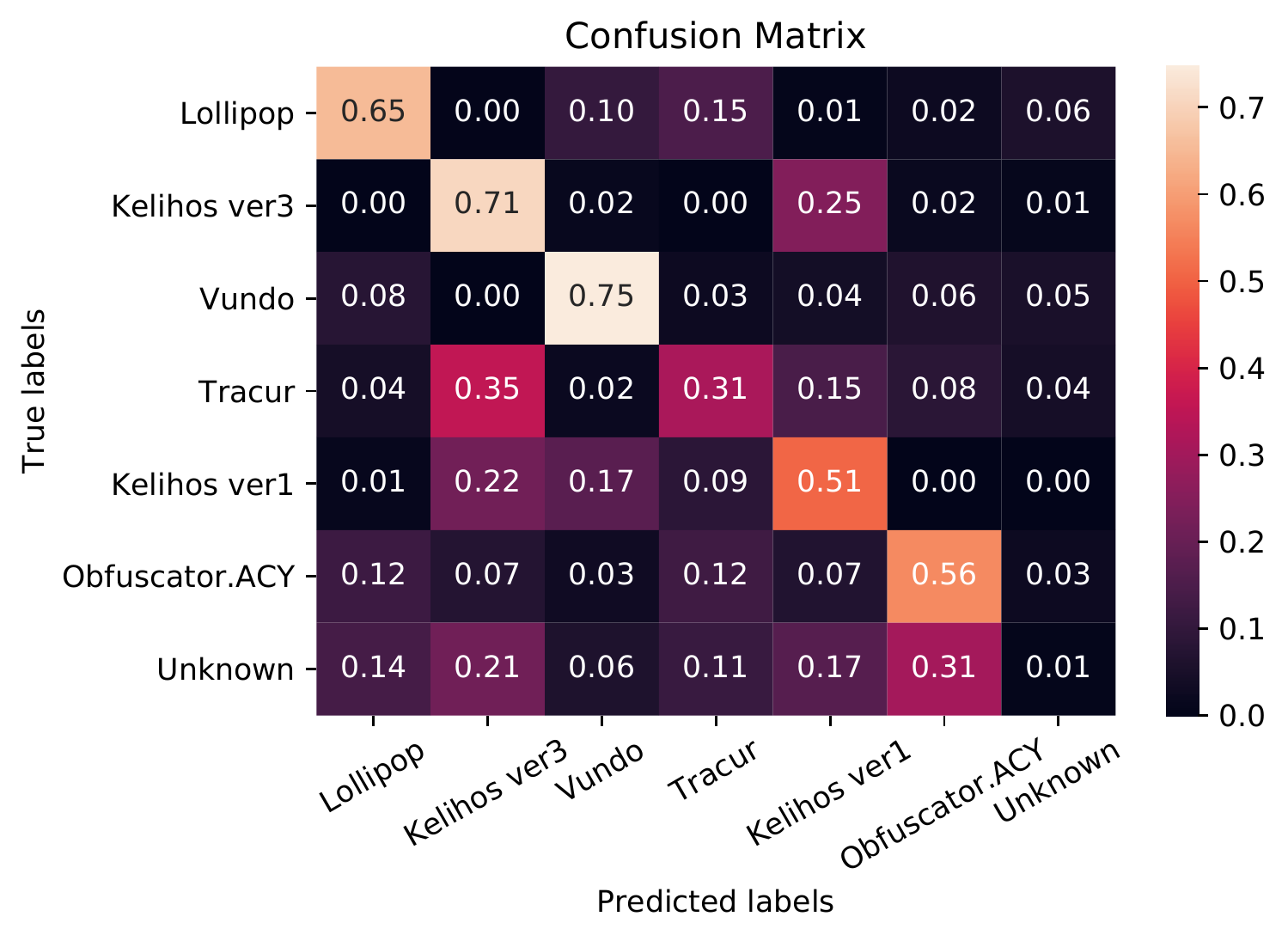}
    \caption{After pre-training (FCG-random)}
\label{fig: hm2}
\end{subfigure}\hfill
\begin{subfigure}[t]{.34\textwidth}
    \includegraphics[width=\linewidth, height=0.2\textheight]{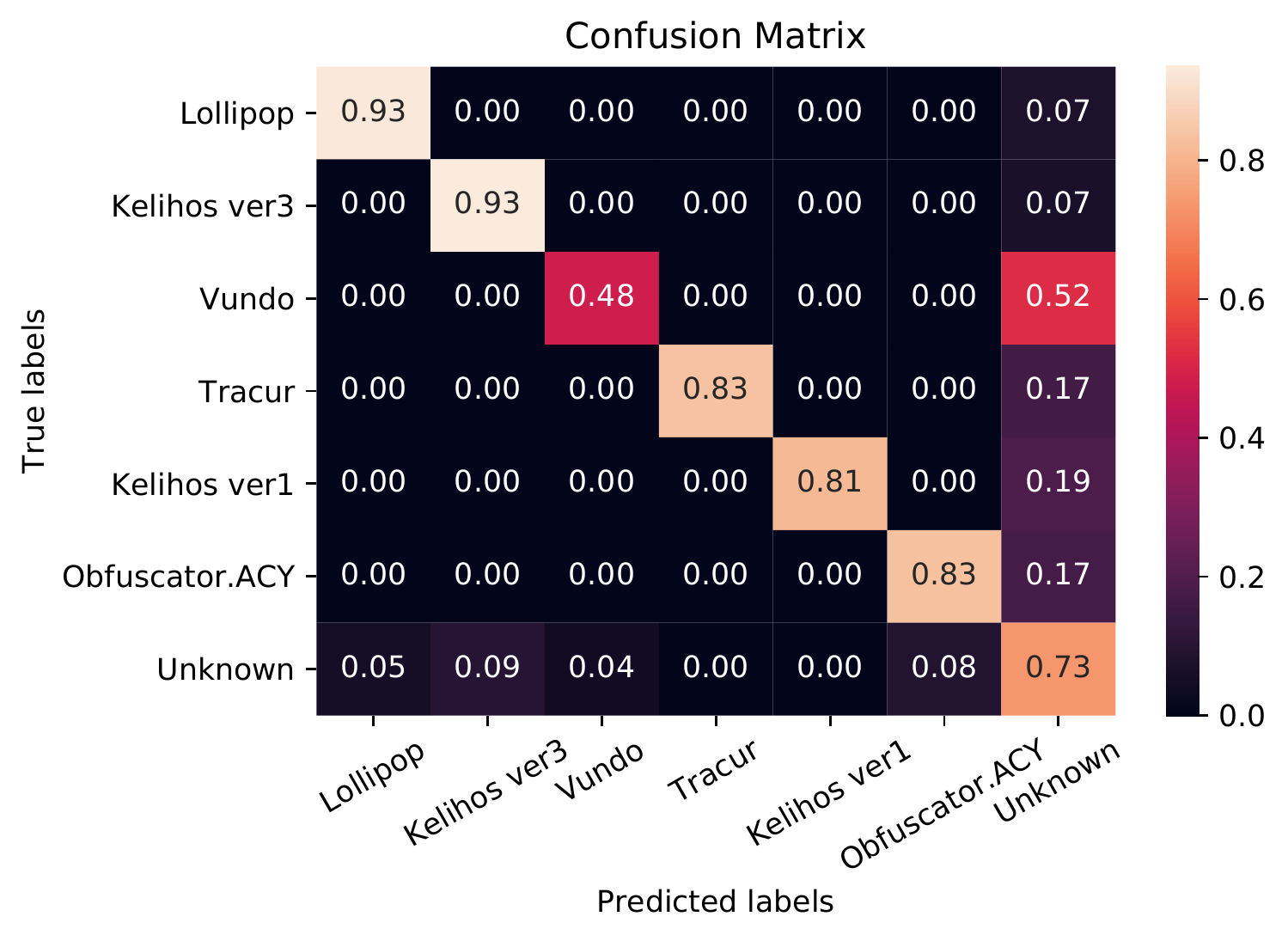}
    \caption{After fine-tuning (FCG-random+ce)}
\label{fig: hm3} 
\end{subfigure}\hfill

\caption{The confusion matirces of the MC test dataset under different settings: (a) Cross-entropy loss without pre-training; (b) Augmented with FCG-random and pre-trained with DTAE; (c) Fine-tuned with cross-entropy loss after (b).}
\label{fig: hm}
\end{figure*}

\begin{figure*}[t]
 \begin{subfigure}[b]{0.49\textwidth}
               \includegraphics[width=\linewidth]{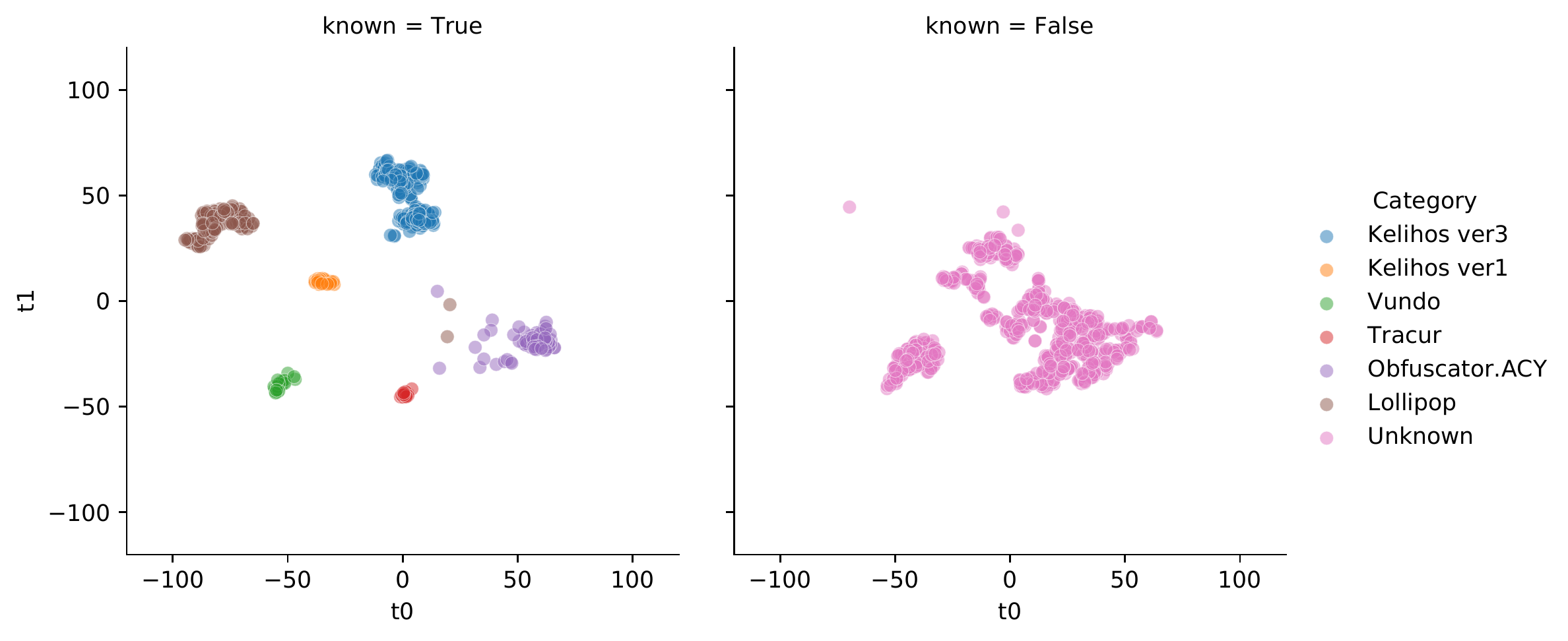}
                \caption{OpenMax}
                \label{fig: tsne8}
        \end{subfigure}
 \begin{subfigure}[b]{0.49\textwidth}               \includegraphics[width=\linewidth]{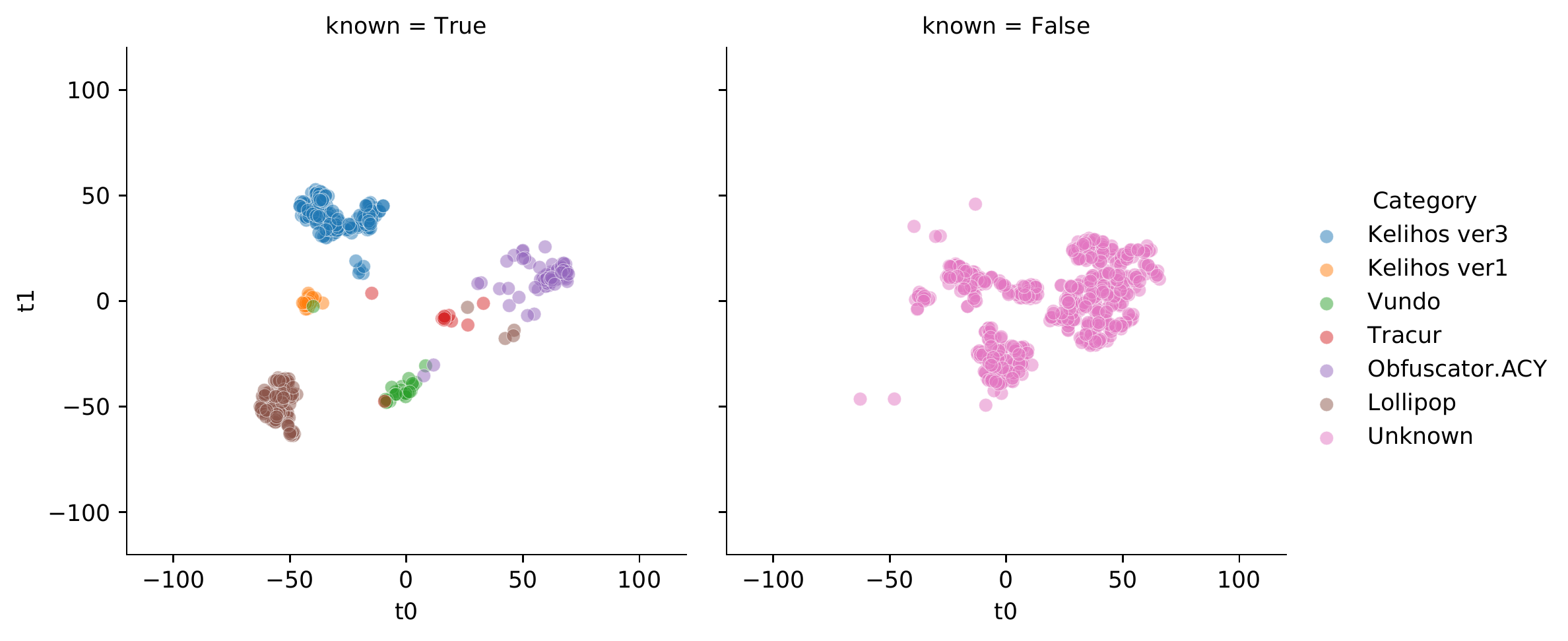}
                \caption{Without pre-training (ce)}
                \label{fig: tsne1}
        \end{subfigure}  \\
 \begin{subfigure}[b]{0.49\textwidth}
               \includegraphics[width=\linewidth]{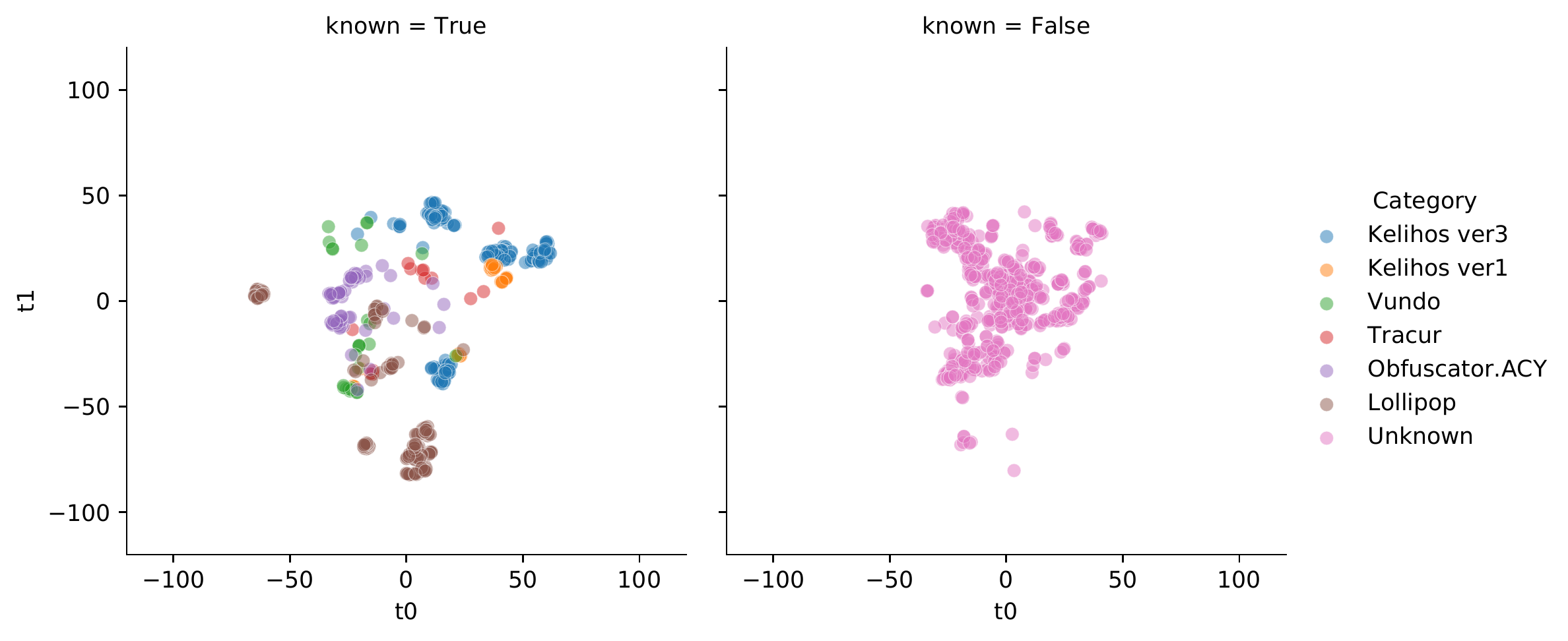}
                \caption{After pre-training (Node dropping)}
                \label{fig: tsne6}
        \end{subfigure} 
  \begin{subfigure}[b]{0.49\textwidth}                \includegraphics[width=\linewidth]{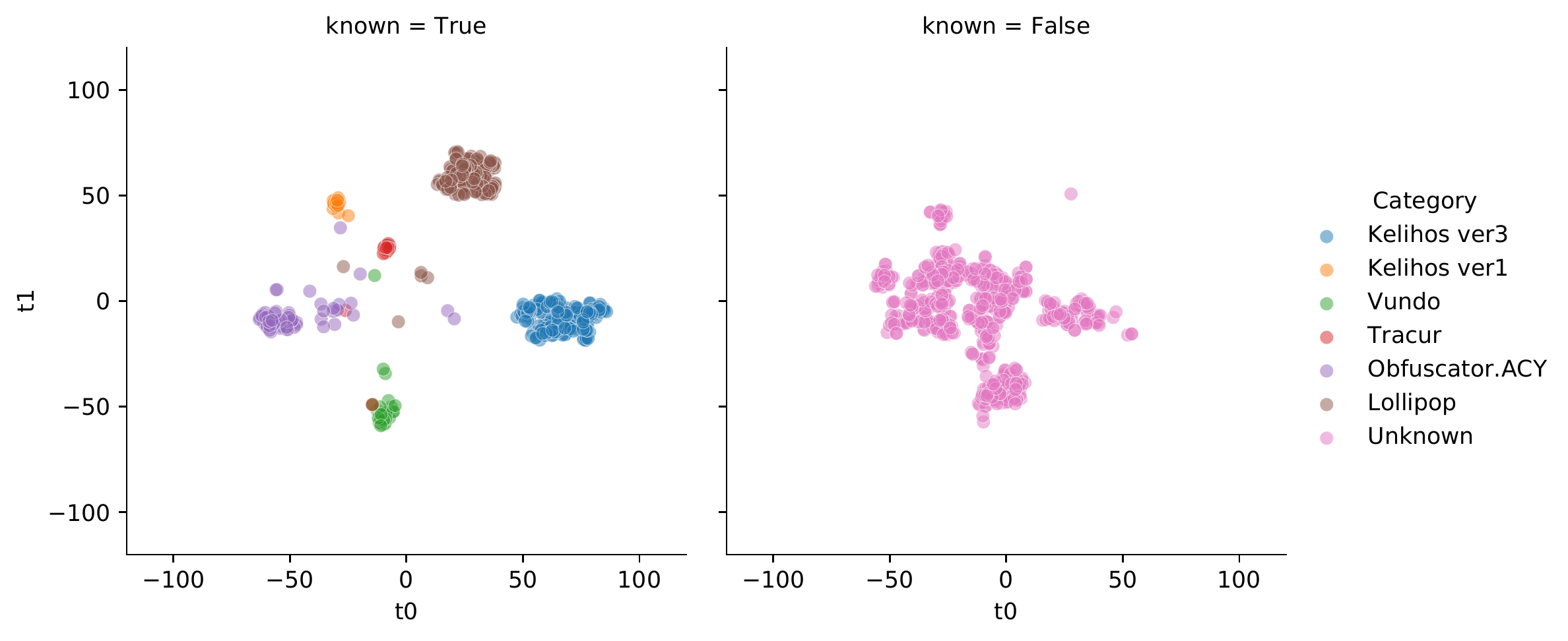}
                \caption{After fine-tuning (Node dropping+ce)}
                \label{fig: tsne7}
        \end{subfigure} \\
 \begin{subfigure}[b]{0.49\textwidth}
               \includegraphics[width=\linewidth]{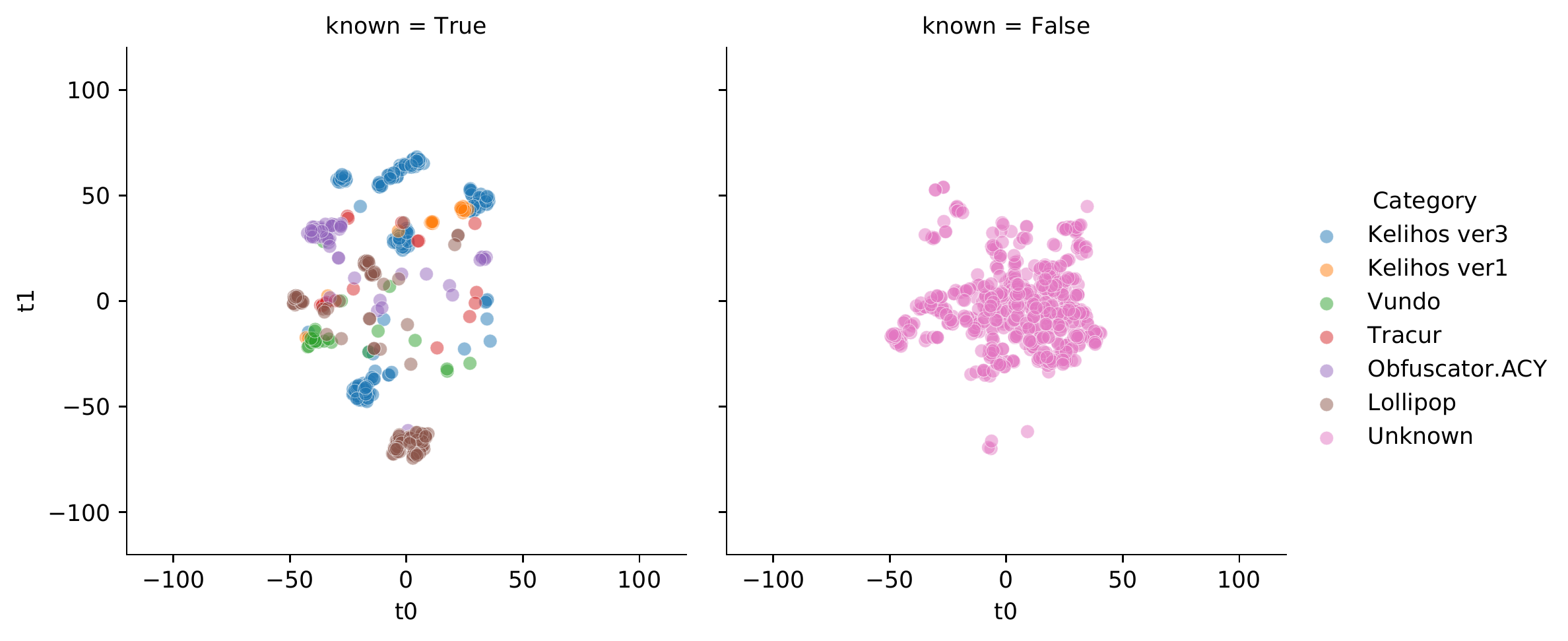}
                \caption{After pre-training (FCG-random)}
                \label{fig: tsne2}
        \end{subfigure} 
  \begin{subfigure}[b]{0.49\textwidth}                \includegraphics[width=\linewidth]{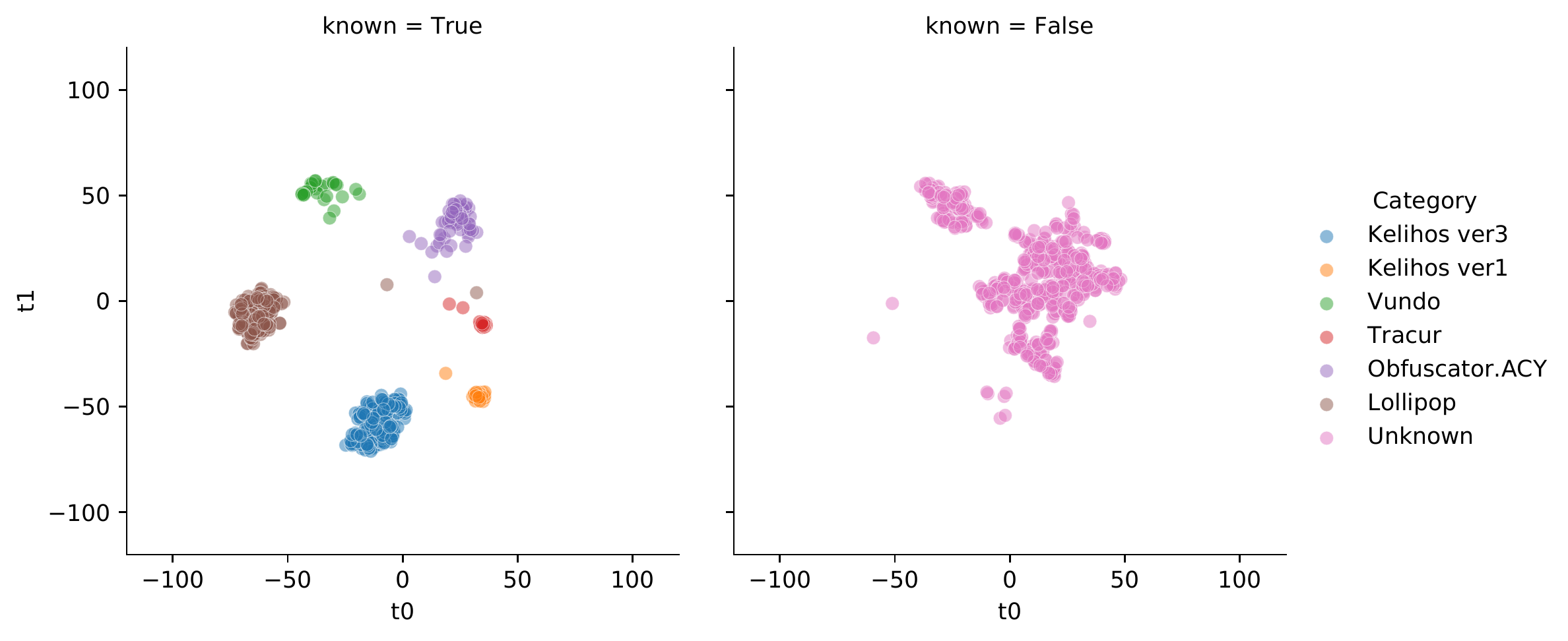}
                \caption{After fine-tuning (FCG-random+ce)}
                \label{fig: tsne3}
        \end{subfigure}

\caption{The t-SNE plots of the MC test representations learned by different settings: (a) OpenMax; (b) Cross-entropy loss without pre-training; (c) Augmented with node dropping and pre-trained with DTAE; (d) fine-tuned with cross-entropy loss after (c); (d) Augmented with FCG-random and pre-trained with DTAE; (e): fine-tuned with cross-entropy loss after (d). The left subplots are the representations of the known class, and the right subplots are the representations of the unknown classes.}
\label{fig: tsne}
\end{figure*}
\subsection{Evaluation Criteria}
To simulate an open-set scenario, we randomly pick six out of nine classes as the known classes and used them in training, and samples from the other classes are regarded as the unknown class, which only exists in the test set. We simulate three different open set groups for each dataset and then repeat each group 10 runs, so each dataset has 30 runs. We calculate the average results of 30 runs for performance evaluation.

We perform a three-dimensional comparison for our proposed approach. First, to show that our proposed approach can achieve good performance in the OSR problem, we compare our proposed approach with the popular OSR solution OpenMax \cite{bendale2016towards}. Moreover, to verify that the self-supervised pre-training process benefits the OSR problem for different downstream loss functions, we compare the model performances with and without using the pre-training process. Finally, we compare our proposed transformation methods ``FCG-shift'' and ``FCG-random'' with other graph transformations ``Node dropping'' (ND) and ``Subgraph sampling'' (SS), which are generally beneficial across datasets \cite{DBLP:conf/nips/YouCSCWS20}. While the AUC score under 100\% FPR is commonly used in model performance measurements, the AUC score under 10\% FPR is more meaningful for malware detection applications. Moreover, we measure the F1 scores for classifying the known classes correctly and recognizing the unknown class correctly for the OSR system. Finally, to show that our proposed statistical approach to recognizing unknown classes in Section \ref{sec: osr} performs as good as the manual thresholding approach: sort the outlier score of the training date in ascending order and then manually pick an outlier score value (99 percentile) as the outlier threshold as in \cite{DBLP:conf/icann/JiaC21,DBLP:conf/sdm/HassenC20,DBLP:journals/corr/abs-2105-13557}, we compare two different thresholding strategies -- ``manual threshold'' and ``statistical threshold'' -- on the representations learned by the vanilla models without pre-training process. To verify that our proposed approaches achieve significant improvement on the OSR, we perform t-tests against OpenMax with
95\% confidence in both the AUC scores and F1 scores.

\subsection{Experimental Results}

We test our proposed pre-training strategy on downstream networks with classification (cross-entropy loss) and representation (triplet loss) loss functions and apply the statistical thresholding approach to learned representations. Table \ref{tab:auc} shows the average ROC AUC scores of the model performances in two malware datasets under different FPR values: 100\% and 10\%. Comparing ``ce'' and ``triplet'' columns with ``OpenMax'' columns, we observe that no matter with or without our proposed pre-training process, the models that use cross-entropy loss and triplet loss perform better than OpenMax for our malware datasets. Furthermore, our proposed pre-training approach outperforms the models without the pre-training process in all 8 cases (2 datasets $ \times $ 2 FPRs $ \times $ 2 loss functions). On the contrary, the DTAE pre-training with node dropping transformation does not benefit the model performance, and the subgraph sampling transformation even hurts the model performance. For MC dataset, the FCG-random transformation works better than the FCG-shift transformation. Meanwhile, their performances differ with different loss functions for the AG dataset.

We also measure the OSR performances via F1 scores under different categories. As shown in Table \ref{tab:f1}. The three categories are: ``Known'', ``Unknown'', and ``Overall''. Specifically, the ``Known'' category is the average F1 scores of the known classes. Moreover, the ``Overall'' category is the average F1 scores of the known and unknown classes. We observe that the pre-training with our proposed transformation methods improves the model performances in the majority of the cases. However, the pre-training with node dropping and subgraph sampling hurts the model performance in most cases. Moreover, the results in the ``manual threshold'' and ``statistical threshold'' rows indicate that our proposed statistical thresholding strategy in Section \ref{sec: osr} can achieve similar performance with the manually selected threshold. Meanwhile, the statistical thresholding approach reduces the number of hyperparameters and alleviates the grid searching process.

Overall, we notice that for both ROC AUC scores and F1 scores, the DTAE pre-training using our proposed transformation approach benefits the model performance in OSR problems. Meanwhile, the transformation method node dropping does not help malware FCGs datasets. As discussed in Section \ref{sec: transformations}, the FCGs are, in general, very sparse graphs. Dropping nodes and subgraph sampling will potentially lose important information about the malware. Meanwhile, our proposed FCG-shift and FCG-random transformation will preserve all the information by creating isomorphic views.

\subsection{Analysis}

While the ROC AUC and F1 scores show that our proposed pre-training approach improves the models' performances, we plot the confusion matrices of one set of the experiments with the MC test set to analyze the experiment results further. In the experiments, the known malware classes are ``Lollipop'', ``Kelihos ver3'', ``Vundo'', ``Tracur'', ``Kelihos ver1'', and ``Obfuscator.ACY'', the remaining three classes together are considered as the unknown class not participating in the training process. Figure \ref{fig: hm1} shows the confusion matrix of the model using cross-entropy without pre-training. Figure \ref{fig: hm2} and Figure \ref{fig: hm3} are the confusion matrices of the model performance after pre-training with FCG-random and after being fine-tuned with cross-entropy loss, respectively. According to the true positive (TP) predictions along the diagonals of the confusion matrices in Figure \ref{fig: hm2}, the model can already classify the known classes after the pre-training stage. Comparing the model performance without pre-training in Figure \ref{fig: hm1} and the one with pre-training in \ref{fig: hm3}, we observe that the TP predictions have been significantly increased for the unknown class. While the TP predictions on the ``Vundo'' class have decreased, the False Positive (FP) predictions (off-diagonal values) happen only between the known classes and the unknown class instead of among the known classes, which indicates that the known classes are more separable.

To visualize the differences between learned representations, we generate the t-SNE plots of the representations at different stages in different experiments as in Figure \ref{fig: tsne}. Specifically, Figure \ref{fig: tsne8} is the t-SNE plot of the learned representations of OpenMax. Figure \ref{fig: tsne1} shows the representations learned by the model using cross-entropy loss without pre-training. Figures \ref{fig: tsne6} and \ref{fig: tsne7} are the representations learned by the model after pre-training with node dropping and being fine-tuned with cross-entropy loss. Figure \ref{fig: tsne2} and Figure \ref{fig: tsne3} are the representations learned by pre-trained model using DTAE with FCG-random and after being fine-tuned with cross-entropy loss. From the left subplot in Figure \ref{fig: tsne6} and Figure \ref{fig: tsne2}, we observe that even without class label information, the self-supervised pre-training model can capture some cluster information. We can find the tiny clusters for the ``Obfuscator.ACY'' class, ``Kelihos ver3'' class and ``Lollipop'' class, which explains the behavior in Figure \ref{fig: tsne6} and Figure \ref{fig: hm2}. Moreover, in Figure \ref{fig: tsne3}, the representations of the known classes in the left subplots are more separate from each other. Meanwhile, the representations of the unknown class are more concentrated near the origin.

\begin{figure}
\resizebox{0.48\textwidth}{!}{%
\centering
\begin{subfigure}[t]{.3 \textwidth}
    \includegraphics[ width=\linewidth]{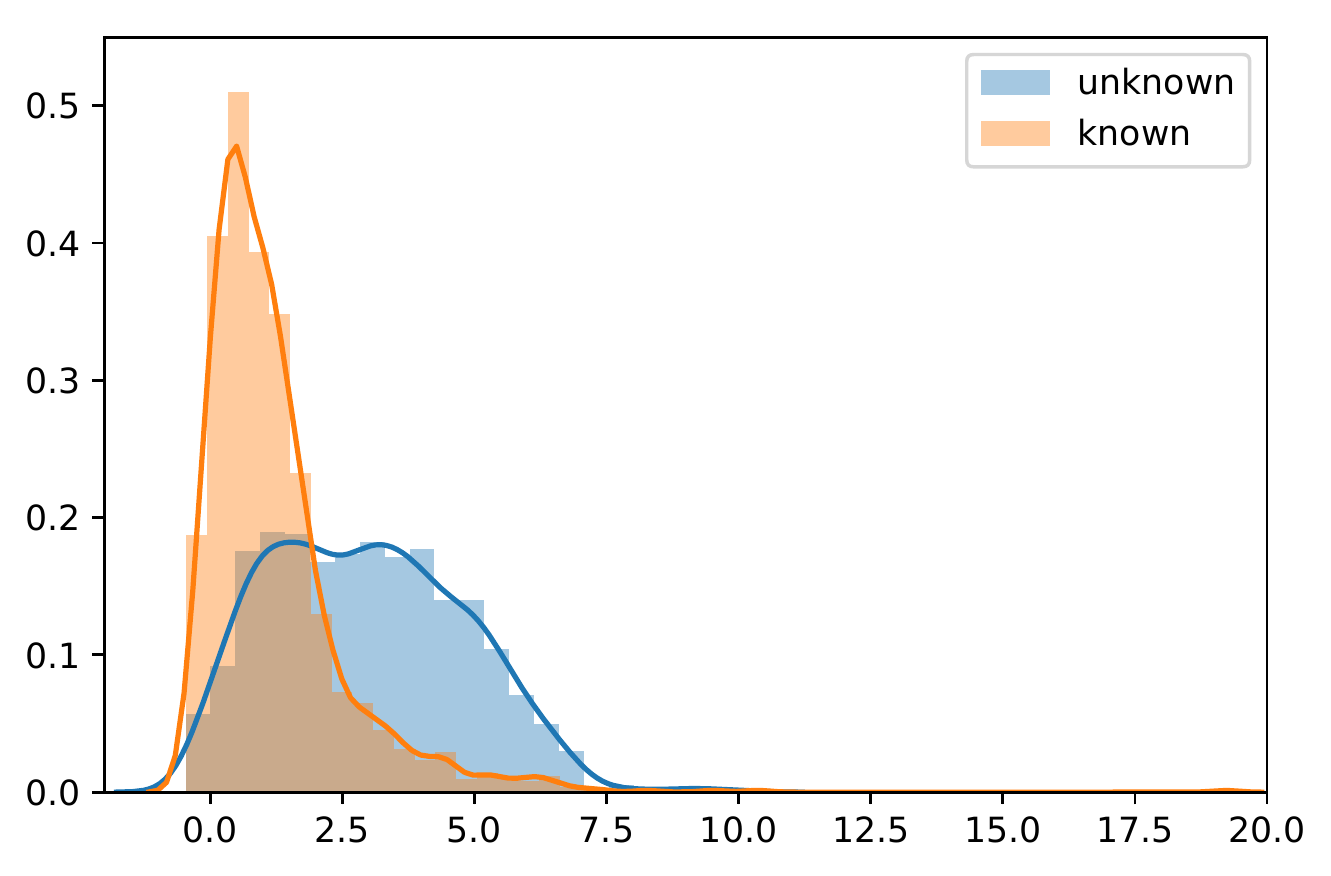}
    \caption{Without pre-training (ce)}
\label{fig: hist1}
\end{subfigure}\hfill
\begin{subfigure}[t]{.3\textwidth}
    \includegraphics[ width=\linewidth]{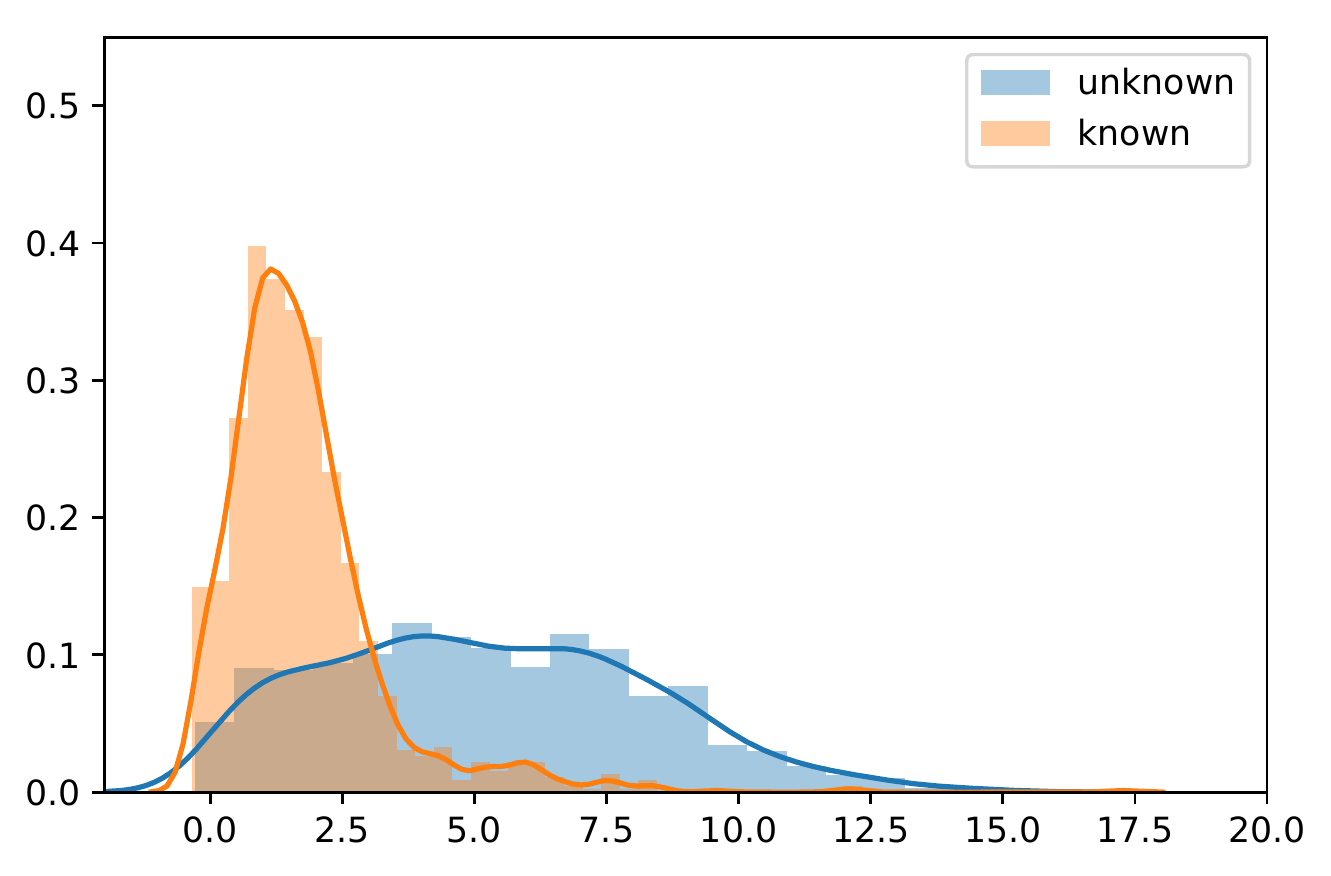}
    \caption{With pre-training (FCG-random + ce)}
\label{fig: hist2}
\end{subfigure}\hfill}
\caption{The distributions of outlier scores for the known and unknown classes of the MC dataset using cross-entropy loss with and without pre-training process.}
\label{fig: hist}
\end{figure}

Figure \ref{fig: hist} shows the distributions of the average outlier scores for the known and unknown classes for the MC test set. Comparing the distributions of outlier scores generated from cross-entropy loss without pre-training in Figure \ref{fig: hist1} and with pre-training in Figure \ref{fig: hist2}, we notice that while the pre-training process increases the outlier scores for both the known classes and the unknown class, it increases the outlier scores in the unknown classes more significantly, which pushes the distribution further away from the known classes. Therefore, there is less overlap and higher accuracy.

\section{Conclusion}
In this paper, we design a two-stage learning process for learning the representations of the malware FCGs to resolve the set recognition problem of malware samples. Specifically, we propose two transformation methods for the FCGs to facilitate the detransformation autoencoder (DTAE) in the pre-training step. Then, we fine-tune the network with different types of loss functions. Moreover, to find the optimal threshold for the OSR problem, we design a statistical thresholding approach based on the distribution of learned representations. The proposed approach reduced the number of hyper-parameters and hence the costs of the resources for the hyperparameter tuning process. We evaluate the pre-training approach with classification loss and representation loss functions on two malware datasets. The results indicate that our proposed approach can improve both model performances for the OSR tasks.

\bibliographystyle{IEEEtran}
\bibliography{root}

\end{document}